\documentclass[12pt,preprint]{aastex}

\newcommand\simlt{\lower.5ex\hbox{$\; \buildrel < \over \sim \;$}}
\newcommand\simgt{\lower.5ex\hbox{$\; \buildrel > \over \sim \;$}}

\slugcomment{To appear in the {\it Astrophysical Journal}}

\begin{document}
\title{Gravitational wave-frequencies and energies in hypernovae}
\author{Maurice H.P.M. van Putten}
\affil{MIT 2-378, Cambridge, MA 02139-4307}

\begin{abstract}
   A torus develops a state of suspended accretion against a magnetic wall around a 
   rapidly rotating black hole formed in core-collapse hypernovae. It hereby
   emits about 10\% of the black hole spin-energy in gravitational 
   radiation from a finite number of multipole mass moments. We quantify
   the relation between the frequency of quadrupole gravitational radiation and 
   the energy output $E_w$ in torus winds by
   $f_{gw}\simeq 470\mbox{Hz}(E_w/4\times 10^{52}\mbox{erg})^{1/2}(7M_\odot/M)^{3/2}$, 
   where $M$ denotes the mass of the black hole. We propose that $E_w$ irradiates the
   remnant stellar envelope from within. We identify $E_w$ with 
   energies $\sim 10^{52}$ erg inferred from X-ray observations on matter injecta;
   and the poloidal curvature in the magnetic wall with the horizon opening angle in baryon poor 
   outflows that power true GRB energies of
   $E_\gamma\simeq 3\times 10^{51}$ erg.
\end{abstract}


\section{Introduction}

  Stellar mass black holes surrounded by a torus of a few tenths of a solar mass
  may form in prompt core-collapse of young massive stars by conservation
  of mass and angular momentum, as in the 
  Woosley-Paczynski-Brown scenario of hypernovae \citep{woo93,pac98,bro00}.
  These are transient systems which are interesting from the point
  of view of the first law of thermodynamics. A torus surrounded a rapidly rotating
  black hole develops a suspended accretion state for the lifetime of 
  rapid spin of the black hole \citep{mvp01} -- generally a secular
  timescale. The torus catalyzes some 10\% of black hole spin-energy into gravitational radiation, 
  with additional winds, thermal and MeV neutrino emissions \citep{mvp02}. Gravitational radiation 
  is emitted by a finite number of mass multiple moments produced by a Papaloizou-Pringle 
  instability
  \citep{pap84,gol86,mvp02c}; torus winds form analogously to pulsar winds, and MeV 
  temperatures develop by internal heating due to shear in response to competing 
  torques acting on the inner and the outer face of the torus. Similar considerations 
  apply to black hole-torus systems formed in black hole-neutron star coalescence, 
  provided that the black hole spins rapidly.

  The predicted gravitational radiation may be determined by upcoming 
  gravitational wave experiments LIGO \citep{abr92}, VIRGO \citep{bra92}, 
  TAMA \citep{tam02} and others, possibly in combination with any of the bar or sphere 
  detectors presently being developed. Accurate prediction of the gravitational 
  wave-frequency is important in estimating the expected signal to noise ratios
  (e.g., \cite{cut02}), or in the design of dedicated experiments, such as
  dual recycling interferometry \citep{har02}. 

  In this {\em Letter}, we quantify
  the relation between the gravitational wave-frequency and the associated energetics
  in torus winds. In a hypernova, these wind energies deposit their momenta onto 
  surrounding remnant envelope, providing enhanced kinetic energies to matter ejecta.
  Wind energies determined from observational data on kinetic energies in GRB-supernova
  events (e.g., \citep{ree02}) and their remnants hereby provide 
  a means for constraining the gravitational wave-frequency, and circumventing uncertainties 
  in model parameters. 
  
  The energetics of black hole-torus systems in suspended accretion depends on 
  the mass of the black hole and the angular velocity of the torus. 
  Hypernovae are believed to
  involve stellar mass black holes of about $4-14M_\odot$, consistent with the 
  observed mass distribution in soft X-ray transients. The angular velocity of 
  defines the efficiency of the catalytic conversion of black hole
  spin-energy as well as the frequency of gravitational radiation
  (at various multipole moments). Yet, it is difficult to determine from first
  principles, and it becomes of interest to consider ways to constrain the 
  angular
  frequency of the torus from the output in torus winds.
  The specific approach presented here 
  considers the the phenomenology of torus winds,
  as well as a detailed calculation on the energy output in gravitational
  radiation, winds and thermal and neutrino emissions.

  The fate of the torus wind energies is determined by its nearby environment.
  We here propose that the torus wind energy is largely deposited within and onto 
  the surrounding remnant stellar envelope, in the form of radiation and kinetic
  energy, respectively. We thus attribute the X-ray line-emission energies in matter 
  ejecta in GRB-supernova
  associated events (e.g., \citep{ree02,ghi02}) to irradiation 
  {\em from within}, as the expanding envelope reaches optical depth of order unity.
  We note that the deposition of torus wind energies is essentially that of a point explosion, 
  given the relatively short durations 
  of activity at hand (about 20 s in association with a long gamma-ray burst). 
  
  It has further been suggested that hypernovae may be associated with supershells 
  \citep{efr98,kim99} which are X-ray bright \citep{wan99}. Indeed, 
  these X-ray bright features appear to be distinct from 
  collective supernovae \citep{bun01}. 
  A baryonic component to torus winds represents ejecta, which may 
  further account
  for chemical enhancements in the companion stars in some of the soft X-ray 
  transients \citep{bro00}, e.g., 
  GRO J1655-40 \citep{isr99} and V4148Sgr \citep{oro02}. 
  
  A GRB-hypernova association is receiving increasing
  support from the association of long gamma-ray bursts with supernovae, as in 
  notably in GRB 980425/1998bw \citep{gal98} and, more recently, GRB 011121 \citep{blo02}, 
  GRB 011211 \citep{ree02} and GRB 020405 \citep{pri02}. Indeed, long GRBs are associated 
  with star-forming regions \citep{blo00}, and hence possibly with young massive stars 
  in binaries. This appears to be consistent with a host environment in molecular 
  clouds \citep{pri02}. In turn, this suggests closer consideration of an association 
  of hypernovae and their remnants in molecular clouds.
  
  Calorimetry on systems consisting of a torus in suspended accretion around a 
  rapidly spinning black hole therefore can be approached as follows.
  First, the beamed output in true GRB energies $E_\gamma$ has been determined to be 
  a few times $10^{50}$ergs \citep{fra01}, which is indicative of ultrarelativistic 
  baryon poor jets of somewhat larger energies. Second, the energy $E_w$ in torus 
  winds can be determined from energies associated with matter ejecta and their remnants
  in the host environment,
  the impact on a binary companion star in a remnant soft X-ray 
  transient, and from beaming in GRBs when derived from collimation by winds. 
  Third, the energy emitted in gravitational radiation of $E_{gw}\simeq 6\times 10^{53}$erg 
  for a ten solar mass extreme Kerr black hole may be determined by upcoming 
  gravitational wave experiments. A fourth energy deposition in MeV neutrinos is
  ulikely to be detectable in the foreseeable future, given the relatively
  large distances of these transient events. The energies $E_w$ and $E_{gw}$
  are both produced by the torus and are a function of its angular velocity. This
  suggests determining $E_w$ in an effort to predict $E_{gw}$ and hence the
  frequency $f_{gw}$ of gravitational radiation, as $E_{gw}$ and $f_{gw}$ are
  closely related. 

  We shall derive simplified expressions for the fractions
  $E_{gw}/E_{rot}$, $E_w/E_{rot}$ and in dissipation $E_{diss}/E_{rot}$,
  relative to the rotational energy $E_{rot}$ of the black hole.
  These results are used to define an observational
  constraint on the torus-to-black hole angular velocity by the energy emitted
  in torus winds. This serves to circumvent otherwise uncertain physical parameters, 
  notably the viscosity in the torus which provides coupling between the inner and 
  the outer face.  
  We propose to determine $E_w$ from
  calorimetry on hypernova ejecta and their remnants. Accurate prediction
  of the gravitational wave spectrum from wind energies will be useful in 
  designing dedicated experiments for searches for gravitational radiation
  from hypernovae.

  \section{Multipole moments in wide tori}

  Quite generally, a torus tends to develop instabilities in response to shear. 
  This can be studied analytically in the approximation of incompressible fluid 
  about an unperturbed angular velocity 
  \begin{eqnarray}
  \Omega=\Omega_T(a/r)^q,
  \label{EQN_ROT}
  \end{eqnarray}
  where $q\ge 3/2$ denotes the rotation index
  and $\Omega_T\simeq M^{1/2}/a^{3/2}$ \citep{pap84}. In the inviscid limit, we are 
  further at liberty to consider the evolution of irrotational modes in response to 
  initially irrotational perturbations to the underlying flow (vortical if $q\ne2$) 
  by Kelvin's theorem. This approach shows the Papaloizou-Pringle instability
  \citep{pap84} to also operate in wide tori \citep{mvp02c}.
  The neutral stability curves of the resulting buckling modes can be described
  in terms of the critical rotation index $q_c=q_c(b/a,m)$ as a function of the
  minor-to-najor radius $b/a$. Quadratic fits to these stability curves are
  \begin{eqnarray}
  q_c(b/a,m)=\left\{
  \begin{array}{rl}
  2.49(b/a)^2+1.73 & (m=2)\\
  6.47(b/a)^2+1.73 & (m=3)\\
  12.4(b/a)^2+1.73 & (m=4)\\
  20.0(b/a)^2+1.73 & (m=5)\\
  0.85m^2(b/a)^2+1.73     & (m>5)
  \end{array}\right.
  \end{eqnarray}
  Instability sets in above these curves, stability below. We note that
  for $m=2$ the critical value $q_c=2$ obtains for 
  \begin{eqnarray}
  b/a=0.3225,
  \end{eqnarray}
  associated with the Rayleigh stability criterion for the azimuthally 
  symmetric wave mode $m=0$. For large $m$, we use the numerical result 
  of critical values $b/a=0.56/m$ for $q=2$.
 
  A quadrupole buckling mode radiates gravitational waves at 
  close to twice the angular frequency of torus \citep{mvp02c}. Because
  the buckling mode represents an internal flow of energy and angular momentum 
  from the inner to the outer face of the torus, in which total energy and angular 
  momentum of the wave remain zero, it is not subject directly to the 
  Chandrasekhar-Friedman-Schutz instability. Nevertheless, it is stimulated
  by gravitational radiation-backreaction forces, as derived in the approximation 
  of the Burke-Thorne potential \citep{mvp02c}.

 \section{Energy output from a torus in suspended accretion}

The suspended accretion state is described by balance of
energy and angular momentum, pertaining to gravitational radiation,
Poynting flux-dominated winds, thermal and neutrino emissions \citep{mvp01}:
\begin{eqnarray}\left\{\begin{array}{l}
        \tau_+=\tau_- + \tau_{rad},\\
\Omega_+\tau_+=\Omega_-\tau_-+\Omega_T\tau_{rad}+P_d,
\end{array}\right.\label{EQN_SAS}\end{eqnarray}
where $P_d$ denotes dissipation in thermal and neutrino emissions,
$\Omega_T=(\Omega_++\Omega_-)/2$ is the angular velocity of the torus,
defined as the mean of the angular velocities of the inner and and outer
faces with angular velocities $\Omega_\pm$. The inner and outer faces
are subject to Maxwell stresses, giving rise to torques $\tau_+$ (ingoing)
and $\tau_-$ (outgoing). In what follows $2\pi A$ shall denote the active 
magnetic flux supported by the torus, representing open magnetic field-lines
connected to infinity and the horizon of the black hole. These
are endowed with no-slip/slip boundary conditions, and thereby assume the
angular velocity of the torus. The remainder of magnetic field-lines are inactive,
making up a toroidal `bag' of closed field-lines, both on the inner and the outer 
face, similar to the closed field-lines in pulsar magnetospheres \citep{gol69,mvp01}.
Thus, $\tau_-=A^2f_w^2\Omega_-$ associated with a fraction $f_w$ of flux in torus 
winds to infinity, and $\tau_+ = (\Omega_H-\Omega_+)A^2f_H^2$ associated with a 
fraction $f_H$ of torus winds entering the black hole. With this definition,
$f_w+f_H=1$. 

We consider shear stresses and the resulting dissipation 
in the torus to be due to turbulent magnetohydrodynamical flow, in response
to competing torques acting on the inner and outer face. 
These competing torques promote super-Keplerian and sub-Keplerian motions,
and hence the torus assumes a geometrically thick shape. 
By dimensional analysis, we consider the constitutive relation
$P_d= \gamma A_r^2(\Omega_+-\Omega_-)^2$ with $\gamma$ a factor
of order unity and $A_r=Rh<B_r^2>^{1/2}$ denoting the root
mean square of the radial flux averaged over the interface between
the two faces with contact area $2\pi Rh$. The detailed structure of the 
magnetohydrodynamical flow which develops determines the ratio 
$\gamma A_r^2/A^2$, which parametrizes the effective viscosity. The 
constitutive law for $P_d$ thereby contains one free parameter.
In what follows,
we parametrize the relative strength of the net radial magnetic field as
\begin{eqnarray}
z=\left(\frac{b}{a}\right)\left(\frac{\gamma A_r^2}{A^2}\right),
\label{EQN_PAR1}
\end{eqnarray}
where $a$ and $b$ denote the major and the minor
radius of the torus, respectively.
This choice of parametrization (\ref{EQN_PAR1}) is such that $z$ becomes
independent of the aspect ratio $b/a$, whenever the magnetohydrodynamical
flow develops a flat infrared spectrum up to the first geometrical break 
$m^*=[a/b]$ in the azimuthal wave-number. 

A leading order expansion in the minor radius $b$ of (\ref{EQN_ROT}) obtains 
$\Omega_\pm=\Omega_T(1\pm\delta)$ with $\Omega_a=\Omega_T$ denoting the mean
angular velocity of the torus and $\Omega_\pm$ the average angular velocity 
of the inner and the outer face. Here,
\begin{eqnarray}
\delta=\frac{qb}{2a}
\label{EQN_PAR2}
\end{eqnarray}
denotes the slenderness ratio of the torus. In particular, we have
$[\Omega]\simeq\Omega_+-\Omega_-\simeq q\Omega_T b/a$. 
The equations (\ref{EQN_SAS}) can be now be solved for 
\begin{eqnarray}
\eta=\frac{\Omega_T}{\Omega_H},
\label{EQN_PAR3}
\end{eqnarray}
and the energy output $E_{gw}$ in gravitational radiation. Explicit expressions for
the energy output $E_w$ and $E_d$ in winds 
and the combined dissipation in heat and neutrino emissions. The model parameters 
are, therefore, the rotation index $q$, the dimensionless magnetohydrodynamical
viscosity $z$, the slenderness ratio $\delta$, the ratio of angular velocities
$\eta$, in addition to flux fractions $f_H$ and $f_w$.

\subsection{Estimate of $\eta$}

The first equation of (\ref{EQN_SAS}) may be used to eliminate $\tau_+$ 
in the second equation, $\Omega\tau_{rad}=\Omega_+\tau_+-\Omega_-\tau_--P_d$,
which obtains
\begin{eqnarray}
P_d=\frac{1}{2}[\Omega]\tau_{rad}+[\Omega]\tau_-.
\label{EQN_P1}
\end{eqnarray}
With the constitutive ansatz for $P_d$ given above, 
it follows that $\tau_{rad}=2A_r^2[\Omega]-2\tau_-.$ The luminosity in gravitational 
becomes $L_{gw}\simeq\Omega_T\tau_{rad}$, in view of the fact that the lower order
multipole moments are essentially in corotation with the torus. Here,
\begin{eqnarray}
\Omega_T\tau_{rad}=\Omega^2A^2\left[2\left(\frac{A_r^2}{A^2}\right)
\left(\frac{[\Omega]}{\Omega_T}\right)-2f_w^2\right]=\alpha\Omega_T^2A^2,
\label{EQN_L1}
\end{eqnarray}
where $\alpha=2qz-2f_w^2$
by $[\Omega]/\Omega_T= qb/a$. With forementioned
expression for $\tau_+$, the first equation of (\ref{EQN_SAS}) with (\ref{EQN_L1}) 
obtains $(\Omega_H-\Omega_+)f_H^2=f_w^2\Omega_- +\alpha\Omega.$
Writing $\Omega_\pm=\Omega_T\pm[\Omega]/2$, we obtain
\begin{eqnarray}
\eta= \frac{f_H^2}{\alpha + f_H^2+f_w^2 + \delta(f_H^2-f_w^2)}.
\label{EQN_AA}
\end{eqnarray}

\subsection{Estimate of $E_{gw}$}
An expression for the luminosity $L_{gw}$ in gravitational radiation may be obtained, 
upon noting that most of the black hole luminosity is irradiated into the torus, i.e.: 
$L_H\simeq \Omega_+\tau_+$. (Output along an open flux-tube by the black hole
in association with a GRB is subdominant; \cite{mvp02}.) We write
$\Omega^2_Tf_H^2=\eta
              \Omega\Omega_Hf_H^2A^2 = \eta
              \left({\Omega_T}/{\Omega_+}\right)
  \Omega_+\tau_+ {\Omega_H}/({\Omega_H-\Omega_+})$
for substitution in (\ref{EQN_L1}) after multiplication by $\alpha/f_H^2$.  
By (\ref{EQN_AA}) with $\Omega_+=\Omega_T(1+\delta)$, we have
$L_{gw}/L_H=\alpha^{-1}f_H^2\eta/(1+\delta-\eta(1+\delta)^2)$, i.e.:
${L_{gw}}/{L_H}\simeq {\alpha}/({\alpha+f_w^2+\delta(\alpha+f_H^2+f_w^2)}),$
upon neglecting terms of the order $\eta\delta$ and higher.
The total rate of conversion of rotational
energy of the black hole, including dissipation in the horizon
\citep{tho86}, is given by
$\dot{E}_H=\Omega_H(\Omega_H-\Omega_+)f_H^2A^2$ over a duration
$T=E_{rot}/\dot{E}_H$. The ratio of energy released in gravitational radiation
as a fraction of the rotational energy $E_{rot}$, therefore, becomes
\begin{eqnarray}
\frac{E_{gw}}{E_{rot}}= 
 \frac{\alpha\eta}{\alpha+f_w^2+\delta(\alpha+f_H^2+f_w^2)}.
\label{EQN_energy}
\end{eqnarray}

\section{Energy estimates for symmetric flux distribution}

Consider a symmetric flux-distribution, given by equal fractions of
open magnetic flux on the inner and the outer face: 
\begin{eqnarray}
f_H=f_w=1/2.
\end{eqnarray}
Writing (\ref{EQN_AA})
in the form of $\eta={f_H^2}/({2qz + (1+\delta)(f_H^2-f_w^2)})$, it reduces to
$\eta={1}/{8qz}.$
A fiducial value of $\eta=0.15$ obtains by considering the Rayleigh 
value $q=2$ and $z=0.56$, which attributes the radial modes in the infrared
magnetohydrodynamical spectrum to an equipartition between the unstable 
Papaloizou-Pringle buckling modes. The estimate (\ref{EQN_energy}) becomes
\begin{eqnarray}
\frac{E_{gw}}{E_{rot}}= \frac{8qz-2}{8qz(8qz(1+\delta/4)-1)}.
\label{EQN_S1}
\end{eqnarray}
The energy released in winds satisfies ${E_w}=\alpha^{-1}f_w^2(1-\delta)^2E_{gw}$, 
whereby 
\begin{eqnarray}
\frac{E_{w}}{E_{rot}}= \frac{(1-\delta)^2}{8qz(8qz(1+\delta/4)-1)}.
\label{EQN_S2}
\end{eqnarray}
Finally, (\ref{EQN_P1}) shows that the energy radiated by dissipation in the torus 
satisfies
\begin{eqnarray}
\frac{E_{d}}{E_{rot}}=\delta\frac{E_{gw}}{E_{rot}}+\frac{2\delta}{1-\delta}
\frac{E_{w}}{E_{rot}}.
\label{EQN_S3}
\end{eqnarray}
The case of strong viscosity (large $z$) and small $\delta$ reduce these expressions
further to the three fractions
\begin{eqnarray}
E_{gw}/E_{rot}\sim \eta,~~~E_w/E_{rot}\sim\eta^2,~~~E_d/E_{rot}\sim\eta\delta.
\label{EQN_S4}
\end{eqnarray}

In association to GRBs,   
we attribute a minor output of black hole-spin energy with the formation of
a baryon poor jet of energy $E_{j}$ along an open magnetic flux-tube of the
black hole. 
A universal horizon half-opening angle defines $E_j\sim \theta_H^4$ to be standard,
and a generally small fraction of rotational energy of the black hole \citep{mvp02}.
The simplest geometrical relationship obtains when
$\theta_H$ is given by
the poloidal curvature of the magnetic field, i.e., 
$\theta_H\simeq 10^o (6M/a)$.
Without fine-tuning, this gives a small fraction $E_j/E_{rot}\sim 10^{-3}$ 
released in baryon poor outflows, accompanied by a spread of one order of magnitude
in response to a spread in torus radius by a factor of two --
in agreement with the observed GRB energies
$E_\gamma$ \citep{fra01}. The associated wind energies 
$E_w/E_{rot}\propto (M/a)^3$ vary somewhat less.


\section{Frequency estimate of quadrupole radiation}

The expressions (\ref{EQN_S4}) provide a link between the energy $E_w$ in torus
winds and the frequency in quadrupolar gravitational radiation, set by the
angular velocity $\eta$ of the torus. We propose to estimate $\eta$ from $E_w$,
to circumvents uncertainties in $qz$. This
gives a frequency of quadrupole gravitational radiation given by
$f_{gw}\simeq {1}/({2\pi M})({E_w}/{E_{rot}})$, or
\begin{eqnarray}
f_{gw}\simeq 
470\mbox{Hz}\left(\frac{E_w}{4\times 10^{52}\mbox{erg}}
 \right)^{1/2}\left(\frac{7M_\odot}{M}\right)^{3/2}.
\label{EQN_FE}
\end{eqnarray} 
The wind-energy scale of $4\times 10^{52}$erg corresponds to $\eta=0.1$ and
$M=7M_\odot$.
We can refine (\ref{EQN_FE}) by including a factor $(1-2\eta)/(1-\delta/2)$, 
in view of (\ref{EQN_S1}) and (\ref{EQN_S2}).
Higher order multipole moments are present for a torus of increasingly
smaller width (see \S2), which emit at commensurably higher
frequencies, and probably at somewhat lower intensities.

\section{Calorimetry on ejecta and hypernova remnants}

Torus wind energies impact the remnant stellar stellar envelope from within.
This results in enhanced kinetic energies in
matter ejecta in GRB-supernova associated events
and, ultimately, impact the host environment.
For GRB 011211 \citep{ree02}, we have
\begin{eqnarray}
E_w\simeq m_{ej}\beta_{ej}\simeq 2\times 10^{52}\left(\frac{\theta_{ej}}{20^o}\right)^2\mbox{erg}
\end{eqnarray}
for ejecta of mass $m_{ej}$ with half-opening angle $\theta_{ej}$ 
and $\beta_{ej}=v_{ej}/c$, where $v_{ej}=0.1c$ 
is the velocity of the ejecta in terms of the velocity of light $c$. 
At present, the angle $\theta_{ej}$ is not constrained observationally. It may
be larger than the GRB beaming angle, in which case we should see an 
over abundance of hypernovae with dim or no GRBs. 
We remark that only 
beamed outflows -- baryon poor from the black hole -- have sufficiently high
luminosiy density per steradian to plow through a stellar envelope
\citep{mac98}.

As the envelope expands, its optical depth reaches order unity and is 
irradiated from within by the accumulated remnant radiation from the torus wind energies.
We attribute the observed X-ray line-emissions to this continuum emission,
whose energies are on the order of $10^{52}$ erg \citep{ghi02}. 

Follow-up calorimetry may be pursued on hypernova remnants, possibly
on X-ray bright supershells. This could be pursued by
comparing this population in a nearby galaxy with
with the true GRB-hypernova event rate of about $5\times 10^{-4}$ times
the supernova rate (\cite{woo98}, including beaming),
and identification of X-ray point sources 
(see \cite{bro00}).

\section{Conclusions}

We quantify the energetics of a torus in suspended accretion against
a magnetic wall around a rapidly rotating black hole formed in hypernovae.
The torus catalyzes most of the black hole-spin energy, radiating a major
fraction in a band limited burst of gravitational radiation. The frequency
sweep is about ten percent during the conversion of 
half the rotational energy of an extreme Kerr black hole.
The durations should agree with  
the de-redshifted distribution of $T_{90}$ of long GRBs (Fig. 1 in
\cite{mvp02c}).

In the slender torus limit, simplified expressions are derived for energies
emitted in gravitational radiation, winds and dissipation in the torus.
We attribute the narrow distribution in observed true gamma-ray energies
to baryon poor jets produced in an open magnetic flux-tube on the black hole
to geometry without fine-tuning: the fraction of rotational energy released
in baryon poor outflows is set by a horizon half-opening angle $\theta_H$ 
given by the poloidal curvature of the magnetic wall.

The frequency of gravitational radiation 
is strongly correlated to the energy output in torus winds by (\ref{EQN_FE}). 
We attribute the energies associated in matter ejecta in GRB-supernova
events to the deposition of torus wind energy within and onto the surrounding
remnant stellar envelope. We propose to hereby constrain the expected frequency
of gravitational radiation by calorimetry on these matter ejecta and their 
remnants in the host environment. The latter
may be pursued by statistics on the kinetic energies in matter ejecta, and 
by calorimetry on hypernova remnants in association with molecular clouds, 
conceivably in the form of X-ray bright supershells. 

{\bf Acknowledgment.} The author thanks B.-C. Koo and S. Kim for 
stimulating discussions on supershells, and the referee for a helpful review. 
This research is supported by NASA Grant 5-7012 and an MIT C.E. Reed Fund.

\end{document}